\documentclass[conference]{IEEEtran}
\IEEEoverridecommandlockouts
\usepackage{amsmath,amssymb,amsfonts}
\usepackage{algorithmic}
\usepackage{graphicx}
\usepackage{textcomp}
\usepackage{xcolor}
\usepackage{siunitx}
\usepackage{overpic}
\usepackage{tikz, pgfplots}
\pgfplotsset{compat=1.18}
\def\BibTeX{{\rm B\kern-.05em{\sc i\kern-.025em b}\kern-.08em
    T\kern-.1667em\lower.7ex\hbox{E}\kern-.125emX}}

\usepackage{eso-pic}
\AddToShipoutPicture*{\footnotesize\sffamily\raisebox{0.75cm}{\hspace{1.65cm}\fbox{\parbox{\textwidth}{\copyright 2024 IEEE.  Personal use of this material is permitted.  Permission from IEEE must be obtained for all other uses, in any current or future media, including reprinting/republishing this material for advertising or promotional purposes, creating new collective works, for resale or redistribution to servers or lists, or reuse of any copyrighted component of this work in other works.}}}}

\begin{document}

\title{Solving Electromagnetic Scattering Problems by Isogeometric Analysis with Deep Operator Learning}

\author{\IEEEauthorblockN{Merle Backmeyer}
\IEEEauthorblockA{
\textit{ETH Zurich}\\
Zurich, Switzerland \\
mbackmeyer@ethz.ch}
\and
\IEEEauthorblockN{Stefan Kurz}
\IEEEauthorblockA{
\textit{ETH Zurich}\\
Zurich, Switzerland \\
stefan.kurz@math.ethz.ch}
\and
\IEEEauthorblockN{Matthias Möller}
\IEEEauthorblockA{
\textit{TU Delft}\\
Delft, the Netherlands \\
m.moller@tudelft.nl}
\and
\IEEEauthorblockN{Sebastian Schöps}
\IEEEauthorblockA{
\textit{TU Darmstadt}\\
Darmstadt, Germany\\
sebastian.schoeps@tu-darmstadt.de}
}

\maketitle

\begin{abstract}
We present a hybrid approach combining isogeometric analysis with deep operator networks to solve electromagnetic scattering problems. The neural network takes a computer-aided design representation as input and predicts the electromagnetic field in a de Rham conforming B-spline basis such that for example the tangential continuity of the electric field is respected. The physical problem is included in the loss function during training. Our numerical results demonstrate that a trained network accurately predicts the electric field, showing convergence to the analytical solution with optimal rate. Additionally, training on a variety of geometries highlights the network's generalization capabilities, achieving small error increases when applied to new geometries not included in the training set.
\end{abstract}

\begin{IEEEkeywords}
integral equations (IEs), computer-aided design
(CAD), non-uniform rational B-spline (NURBS), isogeometric
analysis (IGA), electromagnetic modeling, physics-informed neural networks (PINNs), deep operator networks (DeepONets).
\end{IEEEkeywords}

\section{Introduction}
Solving partial differential equations (PDEs) accurately and efficiently is crucial in computational science, especially for complex physical phenomena such as electromagnetic scattering. One of the established methods for addressing such problems is through integral equations (IEs), particularly the electric field integral equation (EFIE) for scattering problems~\cite{Volakis_2012aa}. It requires a precise discrete representation of the geometry. Isogeometric analysis (IGA) has emerged as a promising method that brings together computer-aided design (CAD) and finite element analysis (FEA) using spline-based functions for both geometry and solutions~\cite{Hughes_2005aa}. The spline spaces in \cite{buffa2011isogeometric,Buffa_2019ac} form a discrete de Rham sequence that allows for consistent field representations in electromagnetic problems on volumes and surfaces. Thus, splines have been established to solve the EFIE numerically, e.g.~\cite{Buffa_2014ab, simpson2018isogemetric, dolz_2019ad, Nolte_2024aa,Hofmann_2024}. While robust and accurate, this method can be computationally intensive, especially for large-scale problems or when multiple evaluations are required, such as in optimization tasks~\cite{Balouchev_2024aa}. 

Physics-informed neural networks (PINNs) are a new approach for solving PDEs that provide rapid post-training evaluations \cite{karniadakis2021physics}. PINNs incorporate physical laws in the loss function during the training process, allowing the neural network to learn the solution to the PDE at discrete collocation points~\cite{raissi2019physics}. When extended to deep operator networks (DeepONets), they have the ability to learn operators, making them applicable to entire problem classes rather than just single problem setups~\cite{wang2021learning}. This capability makes DeepONets particularly attractive for optimization tasks.
However, their heuristic nature and lack of rigorous theoretical foundation limit their reliability and pose challenges in ensuring that the solutions are physically meaningful \cite{longo2022rham}. For example, learning the solenoidal property of a magnetic field can be challenging, which may lead to solutions that do not adhere to fundamental physical laws while it is straightforward to encode this via basis functions. In \cite{markidis_old_2021}, Markidis concludes that while current PINNs are not yet competitive with high-performance computing solvers, hybrid strategies integrating PINNs with traditional approaches show promise for developing new efficient solvers.

Combining IGA’s spline framework with DeepONets’ computational speed, IgANets train neural networks to learn the coefficients of spline basis functions. This hybrid method has shown success for the volumetric discretizations of Poisson equation \cite{moller2021physics}. 
Our work adopts this approach for scattering field problems in the frequency domain represented by a surface integral equation. This poses new challenges, i.e.\ different physics with new function spaces and a solution field that is complex and vector valued. The remainder of the paper is structured as follows. In Section \ref{sec:splines} the discretization with B-splines is described. Section \ref{sec:maxwell} introduces the scattering problem and its discretization in the IGA framework, which is put into a deep neural network framework in Section \ref{sec:DNN}. The numerical experiments in Section \ref{sec:Results} test the implementation with an academic example and show high accuracy of the proposed
method. Finally, we conclude in Section \ref{sec:Conclusion}.

\section{Splines and geometry} 
\label{sec:splines}

Computer-aided design (CAD) models are commonly represented using Non-Uniform Rational B-splines (NURBS) due to their ability to exactly describe conic sections, provide local smoothness control, and intuitively define curves and surfaces~\cite{Cohen_2001aa}. Following standard spline theory as outlined in \cite{Beirao-da-Veiga_2014aa}, we define the B-spline basis with an open knot vector $\Xi = [\xi_1, \ldots, \xi_{k+p+1}] \in [0, 1]^{k+p+1}$, where $k$ denotes the number of control points. The basis functions $b_i^p \in S^p(\Xi)$ for $1 \leq i \leq k$ are defined for $p = 0$ as:
\begin{equation}
b_i^0(\xi) =
\begin{cases} 
1, & \text{if } \xi_i \leq \xi < \xi_{i+1}, \\
0, & \text{otherwise},
\end{cases}
\end{equation}
and for polynomial degree $p > 0$ via the recursive relationship
\begin{equation}
b_i^p(\xi) = \frac{\xi - \xi_i}{\xi_{i+p} - \xi_i} b_i^{p-1}(\xi) + \frac{\xi_{i+p+1} - \xi}{\xi_{i+p+1} - \xi_{i+1}} b_{i+1}^{p-1}(\xi).
\end{equation}

In CAD, boundary representations are typically used, where the overall geometry is given by $\Gamma = \bigcup_{n=1}^{N_\Gamma} \Gamma_n$. In isogeometric analysis, these mappings are represented by several NURBS mappings \cite{Piegl_1997aa}, defined as:
\begin{equation}
\Gamma_n(x, y) = \frac{\sum_{i=1}^{I} \sum_{j=1}^{J} P_{i,j} b_{i}^{p1}(x) b_{j}^{p2}(y) w_{i,j}}{\sum_{k=1}^{I} \sum_{l=1}^{J} b_{k}^{p1}(x) b_{l}^{p2}(y) w_{k,l}}, \quad I,J \in \mathbb{N},
\label{eq: NURBS_surface}
\end{equation}
where $P_{j1,j2}$ and $w_{i1,i2}$ denote the control points and weights, respectively.

\section{Full-wave Maxwell discretization}
\label{sec:maxwell}
Let the free space be denoted by $\Omega$ and the perfectly electric conducting (PEC)  surface
by $\Gamma \subset \Omega$. Then the scattering problem is to
compute the scattered field $\pmb{E}_\mathrm{s}$ for a given incident field $\pmb{E}_\mathrm{i}$ 
that satisfies
\begin{equation}
\left\{\begin{aligned}
\operatorname{curl} \operatorname{curl} \pmb{E}_{\mathrm{s}}-\kappa^2 \pmb{E}_{\mathrm{s}} & =0 \quad &\text { in } \Omega, \\
\pmb{E}_{\mathrm{s}} \times \pmb{n} & =-\pmb{E}_{\mathrm{i}} \times \pmb{n}  \quad &\text { on } \Gamma,
\end{aligned}\right.
   \label{eq: scattering_problem}
\end{equation}
with the wave number $\kappa = \omega \sqrt{\varepsilon \mu}$ defined in terms of
angular frequency $\omega$, permittivity $\varepsilon$ and permeability $\mu$, and
the normal vector $\pmb{n}$ on $\Gamma$. Moreover, we apply the Silver-Müller radiation condition at infinity, see e.g.~\cite[p.~29]{Kurz_2012aa}. Following the approach of Stratton and Chu \cite[Sec.~8.14]{Stratton_1941aa}, the
procedure for solving \eqref{eq: scattering_problem} is to represent the scattered electric
field $E_\mathrm{s}$ by the surface current density $\pmb{J}$.  In particular, $\pmb{J}$ has to fulfill the electric-field integral equation (EFIE), cf. \cite{Buffa_2014ab}
\begin{equation}
    (\mathcal{V} \, {\pmb{J}}) \times \pmb{n}= - \pmb{E}_\mathrm{i} \times \pmb{n},
    \label{eq: EFIE}
\end{equation}
where $\mathcal{V} \,\pmb{J}=\mathcal{V}_{\mathrm{A}} \pmb{J}+ \frac{1}{\kappa^2} \mathcal{V}_{\pmb{\Phi}} \pmb{J}$ contains the vector potential operator 
\begin{equation}
    \mathcal{V}_{\mathrm{A}}  = \int_{\Gamma} g_\kappa(\pmb{x}, \pmb{y}) \pmb{J}(\pmb{y}) \mathrm{d} \Gamma_{\pmb{y}}
\end{equation} and the scalar potential operator 
\begin{equation}
  \mathcal{V}_\mathrm{\pmb{\Phi}} =  \operatorname{grad}_x \int_{\Gamma} g_\kappa(\pmb{x}, \pmb{y})\left(\operatorname{div}_{\Gamma} \pmb{J}\right)(\pmb{y}) \mathrm{d} \Gamma_{\pmb{y}} 
\end{equation}
with the homogeneous space Green's function 
\begin{equation}
    g_\kappa(\pmb{x},\pmb{y}) = \frac{e^{-j\kappa|\pmb{x}-\pmb{y}|}}{4\pi|\pmb{x}-\pmb{y}|}.
\end{equation}

For the numerical solution we discretize the surface current as 
$  \pmb{J}^h = \sum_k j_k \pmb{\varphi}_k,$
with coefficients $j_k$ from $\pmb{j} = \left(j_1,...j_K\right), K\in\mathbb{N}$ and basis functions $\pmb{\varphi}_k\in\pmb{\mathbb{S}}^1_{p}(\Gamma)$.
The spline space $\pmb{\mathbb{S}}^1_{p}(\Gamma)$ is chosen according to a conforming discretization~\cite[Def.~10]{Buffa_2019ac}.
Considering \eqref{eq: EFIE} in a discretized weak formulation, we can state the problem as linear system
\begin{align}
    \mathbf{V} \, \pmb{j} = - \mathbf{f}.
\label{eq: LGS}
\end{align}
Then, by applying $\mathcal{V}$ to $\pmb{J}^h$, we can reconstruct $\pmb{E}_\mathrm{s}^h$. For a detailed derivation of the EFIE in the isogeometric context the reader is referred to the work of Wolf~\cite{wolf2020analysis}.

Note, that all matrix and vector entries and thus the solution of \eqref{eq: LGS} depend on the geometry
which is given by the control points $P_{ij}$ and weights $w_{ij}$ of the spline
surfaces $\Gamma$ in \eqref{eq: NURBS_surface}. 
The linear system of equations in \eqref{eq: LGS} can be solved with (fast) direct or iterative methods, see \cite{Harbrecht_2013ac}. However, for a geometry optimization problem, matrix and vector have to be recomputed and the system solved for each new optimization step which can become computationally prohibitive. The idea in this work is to train a deep neural network to predict the solution vector $\pmb{j}$ representing the surface current density from the geometrical input $G = [P_{i,j}, w_{i,j}]$ for $i = 1, \ldots I, j = 1, \ldots, J$ representing the CAD geometry $\Gamma$.

\section{Deep neural networks}
\label{sec:DNN}

In IgANets, the deep neural network is designed to predict the coefficients of the solution quantity, in this case the surface current density, based on the control points of the geometry. Unlike classical DeepONets, which would directly approximate the solution $\pmb{J}(\pmb{x}_\mathrm{i})$ at discrete points $\pmb{x}_\mathrm{i}$, our DNN outputs the coefficients $j_k$ for the isogeometric basis functions. This ensures that the resulting solution resides within the space spanned by the B-Spline basis functions, maintaining physical plausibility and continuity, such as continuity conditions across interfaces.
For a general introduction into DNNs and specific details for PINNs and DeepONets, the reader is referred to \cite{Goodfellow_2016aa}, \cite{karniadakis2021physics} and \cite{wang2021learning}, respectively. 
The network structure begins with an input layer composed of the geometry parameters $G$. The geometrical information is passed through the net by multiplying with the weights $\theta$ and applying activation functions $\sigma$. The output of the network is the vector of surface current coefficients $\pmb{j}$. The loss function, incorporating the physical constraint, i.e.\ the discrete EFIE, reads
\begin{equation}
    \mathcal{L}(\pmb{j}, G, \theta) = \frac{1}{N} \sum_{n=1}^{N} \left( \left[\mathbf{V}(G) \,  \pmb{j}(G, \theta)\right]_n +  \left[\mathbf{f}(G)\right]_n \right)^2. 
    \label{eq: Loss}
\end{equation}
The weights of the network are adapted using gradient-based optimization algorithms, such as Stochastic Gradient Descent (SGD) or ADAM \cite[Sec.~8.3, 8.5]{Goodfellow_2016aa}. These algorithms adjust the weights by computing the gradient of the loss function with respect to the network parameters $\theta$. Since only the output vector $\pmb{j}$ depends on these weights, the gradient can be efficiently computed using backpropagation, a method implemented in standard DNN libraries such as PyTorch~\cite{PyTorch}. This process ensures that the network minimizes the loss function effectively, improving the accuracy of the predicted coefficients and it is terminated once the loss function values fall below some $\epsilon$. Training geometries are selected to cover a diverse range of shapes, ensuring that the neural network can generalize well across the scenarios needed in the prediction phase later on.

\section{Numerical Results}
\label{sec:Results}
In this section, we apply our theoretical framework to a simple test case with a manufactured solution in order to have an analytical reference solution $\pmb{E}_\mathrm{ref}$ at hand \cite[Sec.~5.2]{wolf2020analysis}. We consider a perfectly electric conducting unit sphere at the origin excited by a Hertzian dipole located at position $\pmb{x}_0 = (0.2, 0.2, 0.2)$ with orientation $\pmb{p}_0 = (0, 0.1, 0.1)$ and wave number $\kappa = 2$.  The electric field of the Hertzian dipole can be found in \cite[p.~411,~(9.18)]{Jackson_1998aa}, which is used to generate the required Dirichlet data. The computed surface quantities then represent the E-field of the dipole in the exterior of the sphere according to the equivalence
theorem \cite[Sec.~7.8]{Balanis_2012aa}. The convergence of the proposed spline-based PINN method is then compared against this analytical solution. Based on suggestions from literature, we use a fully-connected DNN with two layers consisting of 50 neurons each, a sigmoidal activation function and the ADAM optimizer \cite{markidis_old_2021}. Hyperparameters and network architecture have not been optimized for the specific problem. 

\begin{figure}
    \centering
%
%
\definecolor{mycolor1}{rgb}{0.00000,0.44700,0.74100}%
\definecolor{mycolor2}{rgb}{0.85000,0.32500,0.09800}%
\begin{tikzpicture}

\begin{axis}[%
width=0.39\textwidth,
height=5cm,
at={(0.758in,0.573in)},
scale only axis,
x dir=reverse,
xmode=log,
xmin=0.125,
xmax=1,
xtick={0.125,0.25,0.5,1},
xticklabels={{$768$},{$192$},{$48$},{$24$}},
xminorticks=true,
xlabel style={font=\color{white!15!black}},
xlabel={Degrees of freedom (DOFs)},
ymode=log,
ymin=1e-05,
ymax=0.015,
yminorticks=true,
ylabel style={font=\color{white!15!black}},
ylabel={Max pw. error},
axis background/.style={fill=white},
xmajorgrids,
xminorgrids,
ymajorgrids,
yminorgrids,
xlabel near ticks,
ylabel near ticks,
legend style={legend cell align=left, align=left, draw=white!15!black}
]
\addplot [color=mycolor1, mark=o, mark options={solid, mycolor1}]
  table[row sep=crcr]{%
1	0.0143784\\
0.5	0.00135023\\
0.25	0.000180826\\
0.125	2.0124e-05\\
};
\addlegendentry{$p=1$}

\addplot [color=mycolor2, dashed]
  table[row sep=crcr]{%
1	0.015\\
0.5	0.001875\\
0.25	0.000234375\\
0.125	2.9296875e-05\\
};
\addlegendentry{$O(h^3)$}
\end{axis}
\end{tikzpicture}%
    \caption{Maximum pointwise error of the electric field on the unit sphere with wave number $\kappa = 2$. The stopping criterion was $\epsilon \leq 10^{-9}$.}
    \label{fig:convergence}
\end{figure}
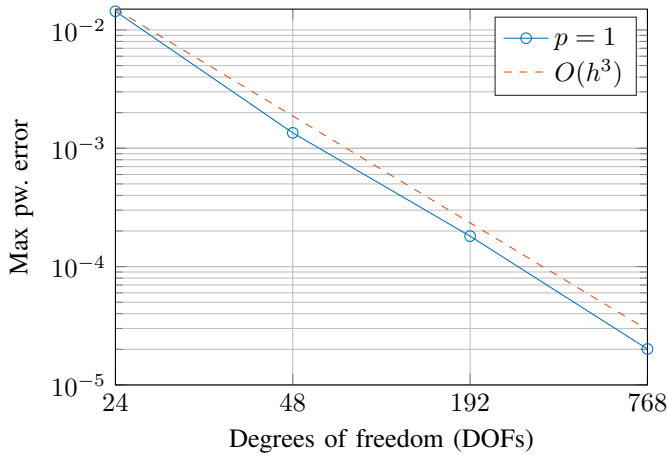

In the first experiment, the network is trained only on a single geometry, i.e., the unit sphere. It is exactly represented by $N_\Gamma=6$ NURBS patches.
The convergence study, shown in Figure~\ref{fig:convergence}, demonstrates that for the basis function space with degree $p=1$, the maximum pointwise error of the computed electric field $\pmb{E_\mathrm{s}^h}$, defined as
\begin{equation}
   \Delta_\mathrm{max} = \max_{\pmb{x} \in S}\left\|\pmb{E}_\mathrm{ref}(\pmb{x}) - \pmb{E}_\mathrm{s}^h(\pmb{x})\right\|_{\mathbb{C}^3},
\end{equation}
where $S$ is a set of points within a radius of $3$ around the geometry, converges with the expected rate of $O(h^3)$ with reference mesh size $h$~\cite[Sec.~5.1]{wolf2020analysis}. For 48 degrees of freedom (DOFs) the error is $\Delta_\mathrm{max} = 1.35\cdot10^{-3}$. Note that for a single geometry, this method reduces to solving \eqref{eq: LGS} via an optimization strategy instead of using a standard linear solver. The real benefit comes from training on multiple geometries.

\begin{figure}
     \centering
     \input{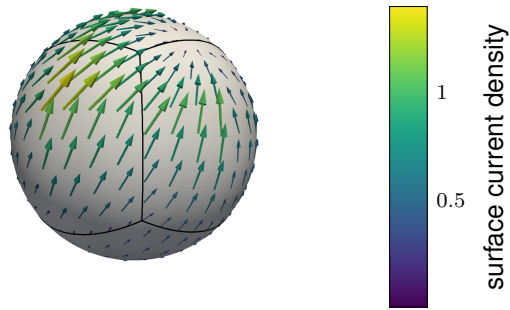}
     \caption{Visualization of the predicted surface current density for an unseen geometry excited by a dipole positioned inside. The stopping criterion was $\epsilon \leq 2\cdot10^{-8}$.}
     \label{fig:surface_current}
 \end{figure}

\cite{Hofmann_2024aa}
So in the next step, we train the network on a range of geometries, transitioning from a sphere to a spheroid, with the main axis radius $r_\mathrm{main}=1$ and the semi-axis radius varies between $r_\mathrm{semi} \in [0.6, 1]$. We consider 100 geometries, with 25\% for training and 75\% for testing. The discretization of the solution field uses basis functions of polynomial degree $p=1$  with $48$ DOFs. The unit sphere is not included in the $25$ training geometries, but used for testing the accuracy of the network's prediction afterwards. The error against the analytical solution is $\Delta_\mathrm{max} =1.37\cdot10^{-3}$, so the difference to training on a single geometry is negligible. The predicted surface current is shown in Figure~\ref{fig:surface_current}.
Evaluating the loss function \eqref{eq: Loss} for each training geometry, the error ranges between $4.29\cdot10^{-8}$ and $2.68\times10^{-6}$, while for testing geometries, it lies between $ 4.28\cdot 10^{-8}$ and $4.15\cdot10^{-6}$, which shows that predictions are not equally accurate for all geometries, but overall the DNN predicts the solution reasonably well.

On a Intel(R) Core(TM) i7-8550U CPU@1.80GHz, the the training process incl. the matrix computations took $\SI{200}{\second}$, afterwards the DNN evaluation takes about $\SI{7}{\milli \second}$ per geometry. Using the linear solver GMRES for \eqref{eq: LGS} instead takes $\SI{0.46}{\second}$ for computing the matrix and solving the system per geometry. Therefore the IGA-PINN approach outperforms the conventional approach if the optimization task requires more than about $400$ evaluations (if the prediction accuracy is acceptable). Please note that we expect significat higher speed-ups for an optimized code with tuned hyperparameters.


\section{Conclusion}
\label{sec:Conclusion}
In this study, we presented IgANets, a novel hybrid approach combining discrete B-spline bases with DeepONets to solve electromagnetic wave problems with an integral equation approach. Leveraging IGA's rigor and DeepONets' efficiency, we trained a neural network to predict solutions for scattered wave problems across various geometries.
A convergence study on a single geometry confirmed the framework's ability to maintain optimal convergence orders. Training on parameterized geometries showed the network's generalization capability with negligible increase of the error. 

This method enables rapid, reliable solution predictions, suitable for optimization and design tasks.

\bibliographystyle{IEEEtran}
\bibliography{bibtex, references}
\end{document}